\documentclass[11pt]{article}
\usepackage[a4paper,margin=1in]{geometry}
\usepackage[T1]{fontenc}
\usepackage[utf8]{inputenc}
\usepackage{lmodern}
\usepackage{amsmath,amssymb,amsthm}
\usepackage{microtype}
\usepackage[hidelinks]{hyperref}

\usepackage{mleftright}
\mleftright

\newtheorem{theorem}{Theorem}
\newtheorem{lemma}[theorem]{Lemma}
\newtheorem{corollary}[theorem]{Corollary}
\newcommand{\Tr}{\operatorname{Tr}}
\newcommand{\E}{\mathbb E}
\newcommand{\Prb}{\Pr}
\newcommand{\ket}[1]{\lvert #1\rangle}

\newcommand{\Dkl}[2]{D\left(#1\middle\|#2\right)}

\usepackage{enumitem}
\usepackage{authblk}

\date{}

\begin{document}
\title{Sharp error tolerance of the monogamy-of-entanglement game}

\author[1,2]{Lloren\c{c} Escol\`a-Farr\`as}
\author[1,2]{Florian Speelman}

\affil[1]{QuSoft, CWI Amsterdam, The Netherlands}
\affil[2]{University of Amsterdam, Amsterdam, The Netherlands}
\renewcommand\Affilfont{\itshape\small}
\maketitle
\begin{abstract}
The BB84 monogamy-of-entanglement (MoE) game has numerous applications
in quantum cryptography, including quantum key distribution, position
verification, and unclonable encryption. In this short note, we improve
the error threshold for parallel repetition with identical answer
strings from the approximately $3.7\%$ certified by the original bound
to the sharp threshold $(1-1/\sqrt2)/2\approx14.64\%$.
When the players need not give identical answers, we also obtain
exponential decay whenever the sum of their error tolerances is below
this threshold. These bounds strengthen existing security reductions
to the BB84 MoE game, improving security bounds and certified noise
tolerances for cryptographic protocols.
\end{abstract}

\section{Introduction}

Monogamy-of-entanglement (MoE) games describe the limits on two
separated players' ability to predict the outcome of a quantum
measurement. The BB84 MoE game, introduced by Tomamichel, Fehr,
Kaniewski, and Wehner~\cite{TFKW13}, was used to prove security
of quantum position verification and one-sided device-independent
quantum key distribution. It also underpins the security analysis
of unclonable encryption introduced by Broadbent and Lord~\cite{BL20},
with further connections developed by Ananth and Kaleoglu~\cite{AK21}.
The wider framework includes monogamy games for subspace coset
states, with applications to unclonable decryption and
copy-protection of pseudorandom functions~\cite{CLLZ21,CV22}.
A leaky variant of the coset-state game has also been used for
unclonable encryption with interactive decryption, unclonable
bit string commitments, and receiver-independent quantum key
distribution~\cite{BC23}. More broadly, MoE games were generalized to extended nonlocal games
by Johnston, Mittal, Russo, and Watrous~\cite{JMRW16}.

In the BB84 MoE game, two separated players prepare a shared state
with the referee, who measures a qubit in a uniformly chosen
computational or Hadamard basis. After learning the basis, both
players must guess the referee's outcome without communicating.
The optimal probability of winning all $n$ parallel repetitions, over all quantum strategies, is
\begin{equation}\label{eq:lambda}
\lambda^n,\qquad\text{where }\quad
\lambda=\frac12+\frac{1}{2\sqrt2}.
\end{equation}
This holds for arbitrary shared states and local measurements
depending on the entire string of basis choices.

For applications that tolerate noise, the players may instead be
allowed a bounded number of errors. We first consider the version
in which they must give \emph{identical answer strings}, with their
common answer differing from the referee's string in at most
$\lfloor\gamma n\rfloor$ positions. This winning condition models
cryptographic settings in which inconsistent answers lead to
rejection, as in position verification.
For this version, the counting argument in Section~III.B
of~\cite{TFKW13} gives
\begin{equation}\label{eq:old-bound}
\Prb[\mathrm{win}]
\leq
\left(\sum_{j=0}^{\lfloor\gamma n\rfloor}\binom nj\right)\lambda^n
\leq \bigl(2^{h_2(\gamma)}\lambda\bigr)^n,
\qquad 0\leq\gamma\leq\tfrac12,
\end{equation}
where $h_2$ is binary entropy with base-two logarithms.
The last expression decays exponentially in $n$ for error rates
$\gamma$ below approximately $3.7\%$.

There is a substantial gap between this certified tolerance and
the error rate of the product Breidbart strategy~\cite{TFKW13}.
In this strategy, the players prepare each referee qubit in a state
halfway between $\ket{0}$ and $\ket{+}$ and both always answer $0$.
For either basis, their answer is incorrect with probability
\begin{equation}\label{eq:threshold}
\gamma_*:=1-\lambda
=\frac12-\frac{1}{2\sqrt2}\approx0.1464.
\end{equation}
Repeating this strategy independently produces identical answer
strings with independent errors. This raises the question:
\emph{is $\gamma_*$ the sharp error threshold under parallel repetition?} In this note, we answer this question affirmatively.
Theorem~\ref{thm:main} replaces the counting bound by
\[
\Prb[\mathrm{win}]\leq 2^{-nD(\gamma\|\gamma_*)},
\qquad 0\leq\gamma<\gamma_*,
\]
where $D$ is binary relative entropy with base-two logarithms.
The product Breidbart strategy attains the same asymptotic exponent
and wins with probability tending to one for $\gamma>\gamma_*$.
Thus both the threshold and the asymptotic exponent are optimal.

The main obstacle to a direct concentration argument is that
the outcomes of an arbitrary parallel strategy need not be
independent. A generalized Chernoff bound due to Panconesi and
Srinivasan~\cite{PS97}, used by Impagliazzo and Kabanets~\cite{IK10}
to derive threshold direct product theorems, replaces
independence by an upper bound on the probability of simultaneous
success on each fixed subset of coordinates. Strong parallel
repetition for the error-free BB84 MoE game supplies precisely
this bound. Combining these two ingredients gives the result.

We also consider players with different error tolerances, without requiring their answers to agree.
Corollary~\ref{cor:unequal-errors} gives exponential decay whenever
the sum of their error tolerances is below $\gamma_*$.
In particular, when one player must guess exactly, the other may have any fixed error tolerance below $\gamma_*$.
This sharpens the full-string guessing estimate used in the one-sided device-independent key-distribution analysis of~\cite{TFKW13}.

Section~\ref{sec:general-games} extends the argument to MoE games
with an error-free parallel repetition bound.
The coset-state and interactive games mentioned above have
different winning conditions, so the BB84 bounds do not directly
apply to them.
Section~\ref{sec:applications} shows how these bounds directly improve
certified error tolerances in position verification and
spacetime-constrained oblivious transfer, and strengthen security bounds for bit commitment and key distribution.

\paragraph{Comparison with arXiv:2611.05943.}
This note was submitted as an arXiv preprint on November 6, 2026.
We had unfortunately overlooked that, while we were finalizing the manuscript, Beigi and Tomamichel independently submitted overlapping results on arXiv two days earlier, which became available November 5~\cite{BT26}.
Their results establish the same sharp error-tolerance threshold for the exponential decay in winning probability of the parallel repetition, and determine the sharp tradeoff between the players' separate error tolerances without requiring identical answers (while our threshold for this case is not tight).
They also present a new resulting key-rate analysis for one-sided device-independent BB84 QKD.
In comparison, our contribution is a shorter and more elementary proof for the identical-answer MoE game.
Additionally, we obtain an explicit finite-size bound achieving the optimal asymptotic decay exponent of its winning probability; the explicit robustness bound of~\cite{BT26} yields a nonoptimal exponent in this setting.

\paragraph{Statement on AI usage.}
The main proof idea was provided by ChatGPT Astra 6. Although the argument is short, the many
applications of the MoE game motivated us to develop it into this
note. We also used ChatGPT to help draft and revise the text.

\section{The BB84 MoE game and its sharp error threshold}

We now define the $n$-fold BB84 MoE game with referee Alice
and players Bob and Charlie, and prove its sharp error threshold.
Bob and Charlie prepare a density operator $\rho_{A^nBC}$, where
$A^n=A_1\cdots A_n$ consists of $n$ qubits and $B,C$ have arbitrary
finite dimensions. They give $A^n$ to Alice and retain $B,C$.
Alice samples $\Theta\in\{0,1\}^n$ uniformly, independently
of the state, and measures qubit $A_i$ in the computational basis
if $\Theta_i=0$, or in the Hadamard basis if $\Theta_i=1$.
Write $X\in\{0,1\}^n$ for the resulting outcome.
After receiving $\Theta$, Bob and Charlie perform local POVMs
$\{P_y^\Theta\}_y$ and $\{Q_z^\Theta\}_z$, producing strings
$Y,Z\in\{0,1\}^n$. They cannot communicate after the preparation
phase. They win if
\begin{equation}\label{eq:winning}
Y=Z\quad\text{and}\quad d_H(X,Y)\leq\lfloor\gamma n\rfloor.
\end{equation}
The agreement condition is checked on the players' outputs; a strategy
may produce different answer strings on some runs.

We denote by $\omega_n(\gamma)$ the supremum of the winning probability
over all such strategies, and by $D(a\|b)$ the binary relative entropy
(Kullback--Leibler divergence) with base-two logarithms:
\[
\Dkl{a}{b}
=a\log_2\frac ab+(1-a)\log_2\frac{1-a}{1-b},
\]
with the endpoint values defined by continuity. 

To control the number of correct answers, we use the following
generalized Chernoff bound. It replaces independence by a bound on
the probability that every variable in a fixed subset equals one.

\begin{lemma}[Generalized Chernoff bound {\cite[Theorem~1.1]{IK10}}]
\label{lem:generalized-chernoff}
Let $J_1,\ldots,J_n$ be $\{0,1\}$-valued random variables and
let $0<p<1$. Suppose that, for every nonempty $S\subseteq[n]$,
\[
\E\left[\prod_{i\in S}J_i\right]\leq p^{|S|}.
\]
Then, for $p<a\leq1$,
\[
\Prb\left[\sum_{i=1}^nJ_i\geq an\right]
\leq2^{-n\Dkl{a}{p}}.
\]
\end{lemma}

\begin{proof}
Put $W=\sum_iJ_i$. For $t\geq0$, the identity
$e^{tJ_i}=1+(e^t-1)J_i$ gives
\[
\begin{aligned}
\E[e^{tW}]
&=1+\sum_{\emptyset\ne S\subseteq[n]}(e^t-1)^{|S|}
\E\left[\prod_{i\in S}J_i\right]\leq1+\sum_{\emptyset\ne S\subseteq[n]}(e^t-1)^{|S|}p^{|S|}
=(1-p+pe^t)^n.
\end{aligned}
\]
The term $1$ corresponds to the empty set. The inequality follows term
by term because all coefficients are nonnegative.
For $p<a<1$, Markov's inequality gives 
\[
\Prb[W\geq an]\leq e^{-tan}\E[e^{tW}]
\leq\bigl[e^{-ta}(1-p+pe^t)\bigr]^n.
\]
Choosing the minimizing value
$t=\ln\bigl(a(1-p)/(p(1-a))\bigr)$ gives
\[
\bigl[e^{-ta}(1-p+pe^t)\bigr]^n=2^{-n\Dkl{a}{p}},
\]
which proves the bound.
For $a=1$, take $t\to\infty$.
\end{proof}

Fix an arbitrary strategy. For each coordinate, let
\[
J_i=\mathbf1\{X_i=Y_i=Z_i\},\qquad W=\sum_{i=1}^nJ_i.
\]
Thus $W$ counts the coordinates on which both players answer correctly
among the $n$ parallel repetitions. The variables $J_i$ need not be
independent.

\begin{lemma}\label{lem:subset}
For every fixed nonempty subset $S\subseteq[n]$,
\begin{equation}\label{eq:subset}
\E\left[\prod_{i\in S}J_i\right]
=\Prb[X_S=Y_S=Z_S]\leq\lambda^{|S|}.
\end{equation}
\end{lemma}

\begin{proof}
Fix a value
$u\in\{0,1\}^{\overline S}$ of the bases outside $S$.
For each choice $\theta_S$ of the remaining bases, retain only the
players' answers on $S$. The corresponding POVM elements are
\[
P_a^{\theta_S,u}=\sum_{y:y_S=a}P_y^{\theta_S,u},
\qquad
Q_a^{\theta_S,u}=\sum_{z:z_S=a}Q_z^{\theta_S,u}.
\]
Let $\rho_S=\Tr_{A_{\overline S}}\rho_{A^nBC}$, and let
$M_a^{\theta_S}$ be Alice's product BB84 measurement projector
on $A_S$ corresponding to outcome $a$ and basis string $\theta_S$.
Summing over Alice's outcomes outside $S$ gives
\begin{equation}\label{eq:reduced-game}
\Prb[X_S=Y_S=Z_S\mid\Theta_{\overline S}=u]
=2^{-|S|}\sum_{\theta_S,a}
\Tr\left[
\bigl(M_a^{\theta_S}\otimes P_a^{\theta_S,u}
\otimes Q_a^{\theta_S,u}\bigr)\rho_S
\right].
\end{equation}
For each fixed $u$, the right-hand side is the success probability of a valid
$|S|$-fold error-free BB84 MoE game. Strong parallel
repetition~\cite{TFKW13} bounds it by $\lambda^{|S|}$.
Since the bases outside $S$ are uniform, it follows that
\[
\begin{aligned}
\Prb[X_S=Y_S=Z_S]
&=2^{-|\overline S|}\sum_{u\in\{0,1\}^{\overline S}}
\Prb[X_S=Y_S=Z_S\mid\Theta_{\overline S}=u]\leq2^{-|\overline S|}\sum_{u\in\{0,1\}^{\overline S}}
\lambda^{|S|}
=\lambda^{|S|}.
\end{aligned}
\]
\end{proof}

We can now combine the subset bound with the generalized Chernoff
bound to prove the main result.

\begin{theorem}\label{thm:main}
Let $n\geq1$ and $0\leq\gamma<\gamma_*$. Every strategy for the
BB84 monogamy-of-entanglement game with winning
condition~\eqref{eq:winning} satisfies
\begin{equation}\label{eq:finite-bound}
\Prb[\mathrm{win}]\leq2^{-n\Dkl{\gamma}{\gamma_*}}.
\end{equation}
\end{theorem}

\begin{proof}
Fix an arbitrary strategy and let $W$ be the number of coordinates on
which both players answer correctly. Lemma~\ref{lem:subset} verifies
the hypothesis of Lemma~\ref{lem:generalized-chernoff} with $p=\lambda$.
Consequently, for $\lambda<a\leq1$,
\begin{equation}\label{eq:chernoff}
\Prb[W\geq an]\leq2^{-n\Dkl{a}{\lambda}}.
\end{equation}

To win the game, the players must give the same answer string and
be correct on at least $n-\lfloor\gamma n\rfloor$ coordinates.
Winning therefore requires $W\geq(1-\gamma)n$. Since
$\gamma<\gamma_*=1-\lambda$, we can apply~\eqref{eq:chernoff} with
$a=1-\gamma$ to obtain
\[
\begin{aligned}
\Prb[\mathrm{win}]
&\leq\Prb[W\geq(1-\gamma)n]\leq2^{-n\Dkl{1-\gamma}{\lambda}}
=2^{-n\Dkl{\gamma}{\gamma_*}}.
\end{aligned}
\]
\end{proof}

The threshold is tight, as shown by the product Breidbart
strategy~\cite{TFKW13}. The players prepare Alice's qubits
in the state $\ket{\beta}^{\otimes n}$, where
\[
\ket{\beta}
=\cos(\pi/8)\ket{0}+\sin(\pi/8)\ket{1},
\]
and both output $0^n$, regardless of the basis string.
Measuring $\ket{\beta}$ in either the computational or Hadamard
basis gives outcome $0$ with probability
$\cos^2(\pi/8)=\lambda$.
The number of errors therefore follows
$\operatorname{Bin}(n,\gamma_*)$.
The winning probability tends to $1/2$ at $\gamma=\gamma_*$
by the central limit theorem, and to one for $\gamma>\gamma_*$
by the law of large numbers.

The same strategy also attains the optimal asymptotic exponent.
For $0\leq\gamma<\gamma_*$, put $\gamma_n=\lfloor\gamma n\rfloor/n$. The binomial
probability of exactly $\gamma_n n$ errors, together with
\eqref{eq:chernoff}, gives
\begin{equation}\label{eq:polynomial-gap}
\frac{2^{-nD(\gamma_n\|\gamma_*)}}{n+1}
\leq\omega_n(\gamma)\leq 2^{-nD(\gamma_n\|\gamma_*)}.
\end{equation}
For the lower bound, use
$\binom nr(r/n)^r(1-r/n)^{n-r}\geq1/(n+1)$ with
$r=\lfloor\gamma n\rfloor$, since $r$ is a mode of
$\operatorname{Bin}(n,r/n)$; the case $r=0$ follows directly.
Taking logarithms and letting $n\to\infty$ yields
\begin{equation}\label{eq:exponent}
\lim_{n\to\infty}-\frac1n\log_2\omega_n(\gamma)
=D(\gamma\|\gamma_*),\qquad 0\leq\gamma<\gamma_*.
\end{equation}

\section{Unequal error tolerances}

Agreement is not needed for the subset bound in
Lemma~\ref{lem:subset}, or for the resulting tail bound on $W$.
We can therefore also allow different error tolerances for the two
players, as in Section~III.B of~\cite{TFKW13}.

\begin{corollary}\label{cor:unequal-errors}
Let $n\geq1$ and $\gamma_B,\gamma_C\geq0$ satisfy
$\gamma_B+\gamma_C<\gamma_*$. Consider the BB84 monogamy-of-entanglement
game in which the players win if
\[
d_H(X,Y)\leq\lfloor\gamma_B n\rfloor
\qquad\text{and}\qquad
d_H(X,Z)\leq\lfloor\gamma_C n\rfloor,
\]
without requiring agreement between $Y$ and $Z$. Every strategy satisfies
\begin{equation}\label{eq:unequal-errors}
\Prb[\mathrm{win}]\leq
2^{-n\Dkl{\gamma_B+\gamma_C}{\gamma_*}}.
\end{equation}
\end{corollary}

\begin{proof}
A coordinate fails to be jointly correct only if at least one player
answers it incorrectly. Thus, whenever the players win this game,
\[
n-W\leq d_H(X,Y)+d_H(X,Z)
\leq(\gamma_B+\gamma_C)n.
\]
Apply~\eqref{eq:chernoff} with $a=1-\gamma_B-\gamma_C$ and use
$D(1-\gamma_B-\gamma_C\|\lambda)
=D(\gamma_B+\gamma_C\|\gamma_*)$.
\end{proof}

For equal tolerances $\gamma_B=\gamma_C=\gamma$, this gives
exponential decay whenever $\gamma<\gamma_*/2\approx7.32\%$.
Unlike Theorem~\ref{thm:main}, this corollary does not establish
the sharp threshold or optimal exponent for the unequal-error game.

In the one-sided device-independent key-distribution proof
of~\cite{TFKW13}, Bob's answer may contain errors, whereas Eve
must guess the referee's entire string correctly. Taking
$\gamma_B=\gamma$ and $\gamma_C=0$ in
Corollary~\ref{cor:unequal-errors} gives
\begin{equation}\label{eq:qkd-guessing}
\Prb\left[d_H(X,Y)\leq\lfloor\gamma n\rfloor,\ Z=X\right]
\leq2^{-n\Dkl{\gamma}{\gamma_*}},
\qquad 0\leq\gamma<\gamma_*.
\end{equation}

\section{Extension to general games}
\label{sec:general-games}

Let $\mathsf G$ be a monogamy-of-entanglement game in which both
players receive the referee's question and must guess its outcome.
Parallel repetition uses independent questions and product referee
measurements, while allowing arbitrary shared states and local
measurements depending on all questions. Write
$\omega^*(\mathsf G^{\times s})$ for the optimal probability of winning
all $s$ coordinates.

\begin{theorem}\label{thm:general-errors}
Suppose that some $0<p<1$ satisfies, for every integer $s\geq1$,
\begin{equation}\label{eq:general-parallel-assumption}
\omega^*(\mathsf G^{\times s})\leq p^s.
\end{equation}
For any strategy in $\mathsf G^{\times n}$, let $E_n$ count the
coordinates on which at least one player answers incorrectly.
Then, for $0\leq\gamma<1-p$,
\begin{equation}\label{eq:general-error-bound}
\Prb[E_n\leq\lfloor\gamma n\rfloor]
\leq2^{-n\Dkl{\gamma}{1-p}}.
\end{equation}
\end{theorem}

\begin{proof}
Fixing the questions outside a subset $S$ and retaining only the
answers on $S$ gives a valid strategy for
$\mathsf G^{\times|S|}$, as in Lemma~\ref{lem:subset}.
Thus the indicators $J_i$ of joint correctness satisfy
$\E[\prod_{i\in S}J_i]\leq p^{|S|}$.
Apply Lemma~\ref{lem:generalized-chernoff} with $a=1-\gamma$,
using $E_n=n-\sum_iJ_i$ and $D(1-\gamma\|p)=D(\gamma\|1-p)$.
\end{proof}

\section{Applications}
\label{sec:applications}

The above bounds sharpen several error-tolerant BB84 MoE security
reductions. Other assumptions and terms in the cited analyses remain
unchanged.

\begin{itemize}[leftmargin=*,itemsep=3pt,parsep=0pt]
\item \emph{Position verification.}
For the position-verification protocol in the random-oracle
model~\cite{Unruh14}, Theorem~\ref{thm:main} raises the certified
error tolerance from approximately $3.7\%$ to every fixed
$\gamma<\gamma_*\approx14.64\%$.
The random-oracle-query terms remain unchanged.

\item \emph{Parallel position verification.}
The above results increase the certified error tolerance of the
BB84-based quantum position-verification protocols
of~\cite{EFS25,EFSS26} from approximately $3.7\%$ to every fixed
$\gamma<\gamma_*\approx14.64\%$.
For~\cite{EFS25}, replace the monogamy factor $\lambda_\gamma$
throughout the analysis by $2^{-D(\gamma\|\gamma_*)}$, including
in the resource condition and probability guarantee over the
choice of the protocol function $f$.
For~\cite{EFSS26}, these conditions remain unchanged, and applying
Theorem~\ref{thm:main} to each committed subset gives
$\Prb[\mathrm{accept}]\leq3\cdot2^{-kD(\gamma\|\gamma_*)}$,
where $k$ is the minimum number of committed positions.

\item \emph{Spacetime-constrained oblivious transfer.}
Corollary~\ref{cor:unequal-errors} replaces
$(2^{2h_2(\gamma)}\lambda)^n$ in~\cite[Appendix~B]{PG16}
by $2^{-nD(2\gamma\|\gamma_*)}$ when each recipient may make
$\gamma n$ errors, without requiring agreement.
The certified tolerance increases from approximately $1.5\%$
to every fixed $\gamma<\gamma_*/2\approx7.32\%$.
The same holds for~\cite[Section~IV]{PG19}, with a union bound
over pairs of output regions.

For~\cite{PGK18}, Eq.~(5) bounds joint correctness on any fixed
subset $S$ by $\lambda^{|S|}$: pad the other positions with
independent encoding pairs, fix their bases, and retain the
attack's choice bit without postselection.
Lemma~\ref{lem:generalized-chernoff} therefore gives the same
improvement to Eq.~(25), despite recipients guessing different
strings. These tolerances need not be tight.

\item \emph{Relativistic bit commitment.}
For~\cite[Section~VI]{PGK18}, let $p_b$ be the probability of
successfully opening bit $b$ under a fixed cheating commitment
strategy, allowing $\gamma n$ errors per opening.
The spacelike separation of the tests and the noisy SCOT bound give
\[
p_0+p_1\leq1+2^{-nD(2\gamma\|\gamma_*)},
\qquad 0\leq\gamma<\gamma_*/2.
\]

\item \emph{One-sided device-independent key distribution.}
Equation~\eqref{eq:qkd-guessing} sharpens the guessing estimate
in~\cite[Eq.~(11)]{TFKW13} and hence the secrecy bound.
Parameter estimation, error-correction leakage, and privacy
amplification still enter; $14.64\%$ is not a key-rate threshold.
\end{itemize}

\bibliographystyle{alpha}
\bibliography{references}

@article{TFKW13,
  author = {Tomamichel, Marco and Fehr, Serge and Kaniewski, J{\k{e}}drzej and Wehner, Stephanie},
  title = {A monogamy-of-entanglement game with applications to device-independent quantum cryptography},
  journal = {New Journal of Physics},
  volume = {15},
  pages = {103002},
  year = {2013},
  doi = {10.1088/1367-2630/15/10/103002},
  note = {\href{https://arxiv.org/abs/1210.4359}{arXiv:1210.4359}}
}

@article{PS97,
  author = {Panconesi, Alessandro and Srinivasan, Aravind},
  title = {Randomized distributed edge coloring via an extension of the {Chernoff--Hoeffding} bounds},
  journal = {SIAM Journal on Computing},
  volume = {26},
  number = {2},
  pages = {350--368},
  year = {1997},
  doi = {10.1137/S0097539793250767},
  note = {\href{https://doi.org/10.1137/S0097539793250767}{doi:10.1137/S0097539793250767}}
}

@inproceedings{IK10,
  author = {Impagliazzo, Russell and Kabanets, Valentine},
  title = {Constructive proofs of concentration bounds},
  booktitle = {Approximation, Randomization, and Combinatorial Optimization. Algorithms and Techniques ({APPROX/RANDOM} 2010)},
  pages = {617--631},
  year = {2010},
  note = {Full version: \href{https://www2.cs.sfu.ca/~kabanets/papers/RANDOM2010.pdf}{authors' manuscript}}
}

@inproceedings{BL20,
  author = {Broadbent, Anne and Lord, S{\'e}bastien},
  title = {Uncloneable quantum encryption via oracles},
  booktitle = {15th Conference on the Theory of Quantum Computation, Communication and Cryptography ({TQC} 2020)},
  series = {Leibniz International Proceedings in Informatics (LIPIcs)},
  volume = {158},
  pages = {4:1--4:22},
  year = {2020},
  doi = {10.4230/LIPIcs.TQC.2020.4},
  note = {\href{https://doi.org/10.4230/LIPIcs.TQC.2020.4}{doi:10.4230/LIPIcs.TQC.2020.4}}
}

@misc{AK21,
  author = {Ananth, Prabhanjan and Kaleoglu, Fatih},
  title = {Unclonable encryption, revisited},
  year = {2021},
  eprint = {2103.15009},
  archivePrefix = {arXiv},
  note = {\href{https://arxiv.org/abs/2103.15009}{arXiv:2103.15009}}
}

@misc{BT26,
  title={Norm Bounds on Sums of Positive Operators with Applications to Monogamy Games and Quantum Key Distribution},
  author={Beigi, Salman and Tomamichel, Marco},
  eprint = {2611.05943},
  archivePrefix = {arXiv},
  year={2026},
  note = {\href{https://arxiv.org/abs/2611.05943}{arXiv:2611.05943}}
}

@misc{CLLZ21,
  author = {Coladangelo, Andrea and Liu, Jiahui and Liu, Qipeng and Zhandry, Mark},
  title = {Hidden cosets and applications to unclonable cryptography},
  year = {2021},
  eprint = {2107.05692},
  archivePrefix = {arXiv},
  note = {\href{https://arxiv.org/abs/2107.05692}{arXiv:2107.05692}}
}

@article{CV22,
  author = {Culf, Eric and Vidick, Thomas},
  title = {A monogamy-of-entanglement game for subspace coset states},
  journal = {Quantum},
  volume = {6},
  pages = {791},
  year = {2022},
  doi = {10.22331/q-2022-09-01-791},
  note = {\href{https://doi.org/10.22331/q-2022-09-01-791}{doi:10.22331/q-2022-09-01-791}}
}

@misc{BC23,
  author = {Broadbent, Anne and Culf, Eric},
  title = {Uncloneable cryptographic primitives with interaction},
  year = {2023},
  eprint = {2303.00048},
  archivePrefix = {arXiv},
  note = {\href{https://arxiv.org/abs/2303.00048}{arXiv:2303.00048}}
}

@inproceedings{Unruh14,
  author = {Unruh, Dominique},
  title = {Quantum position verification in the random oracle model},
  booktitle = {{CRYPTO} 2014},
  year = {2014},
  note = {Full version: \href{https://eprint.iacr.org/2014/118}{Cryptology ePrint Archive, Paper 2014/118}}
}

@misc{EFS25,
  author = {Escol{\`a}-Farr{\`as}, Lloren{\c{c}} and Speelman, Florian},
  title = {Quantum position verification in one shot: parallel repetition of the {$f$-BB84} and {$f$-routing} protocols},
  year = {2025},
  eprint = {2503.09544v1},
  archivePrefix = {arXiv},
  note = {\href{https://arxiv.org/abs/2503.09544v1}{arXiv:2503.09544v1}}
}

@misc{EFSS26,
  author = {Escol{\`a}-Farr{\`as}, Lloren{\c{c}} and {\v{S}}kori{\'c}, Boris and Speelman, Florian},
  title = {Arbitrarily loss-tolerant quantum position verification in a single execution},
  year = {2026},
  eprint = {2606.25037v2},
  archivePrefix = {arXiv},
  note = {\href{https://arxiv.org/abs/2606.25037v2}{arXiv:2606.25037v2}}
}

@article{PG16,
  author = {Pital{\'u}a-Garc{\'i}a, Dami{\'a}n},
  title = {Spacetime-constrained oblivious transfer},
  journal = {Physical Review A},
  volume = {93},
  pages = {062346},
  year = {2016},
  doi = {10.1103/PhysRevA.93.062346},
  note = {\href{https://doi.org/10.1103/PhysRevA.93.062346}{doi:10.1103/PhysRevA.93.062346}}
}

@article{JMRW16,
  author  = {Johnston, Nathaniel and Mittal, Rajat and
             Russo, Vincent and Watrous, John},
  title   = {Extended non-local games and monogamy-of-entanglement games},
  journal = {Proceedings of the Royal Society A},
  volume  = {472},
  number  = {2189},
  pages   = {20160003},
  year    = {2016},
  doi     = {10.1098/rspa.2016.0003},
  note    = {\href{https://doi.org/10.1098/rspa.2016.0003}
                 {doi:10.1098/rspa.2016.0003}}
}

@article{PGK18,
  author = {Pital{\'u}a-Garc{\'i}a, Dami{\'a}n and Kerenidis, Iordanis},
  title = {Practical and unconditionally secure spacetime-constrained oblivious transfer},
  journal = {Physical Review A},
  volume = {98},
  pages = {032327},
  year = {2018},
  note = {\href{https://arxiv.org/abs/1811.07410}{arXiv:1811.07410}}
}

@article{PG19,
  author = {Pital{\'u}a-Garc{\'i}a, Dami{\'a}n},
  title = {One-out-of-{$m$} spacetime-constrained oblivious transfer},
  journal = {Physical Review A},
  volume = {100},
  pages = {012302},
  year = {2019},
  doi = {10.1103/PhysRevA.100.012302},
  note = {\href{https://doi.org/10.1103/PhysRevA.100.012302}{doi:10.1103/PhysRevA.100.012302}}
}
\end{document}